\documentclass[journal,twoside,web]{ieeecolor}
\usepackage{tmi}
\usepackage{cite}
\usepackage{amsmath,amssymb,amsfonts}
\usepackage{algorithmic}
\usepackage{graphicx}
\usepackage{textcomp}
\usepackage{ amssymb }
\usepackage{graphicx}
\usepackage{soul,color}
\usepackage{chapterbib}
\usepackage{ amssymb }
\usepackage{graphicx}
\usepackage{soul,color}
\usepackage{multirow}
\usepackage{hyperref}
\usepackage{comment}

\usepackage[table,xcdraw]{xcolor}
\usepackage[table,xcdraw]{xcolor}
\def\BibTeX{{\rm B\kern-.05em{\sc i\kern-.025em b}\kern-.08em
    T\kern-.1667em\lower.7ex\hbox{E}\kern-.125emX}}

\begin{document}

\title{KiU-Net: Overcomplete Convolutional Architectures for Biomedical Image and Volumetric Segmentation }
\author{Jeya Maria Jose Valanarasu, \IEEEmembership{Student Member, IEEE}, Vishwanath A. Sindagi, \IEEEmembership{Student Member, IEEE}, Ilker Hacihaliloglu, \IEEEmembership{Member, IEEE}, and Vishal M. Patel, \IEEEmembership{Senior Member, IEEE}
	\thanks{Jeya Maria Jose Valanarasu, Vishwanath A. Sindagi and Vishal M. Patel are with the Whiting School of Engineering, Johns Hopkins University, 3400 North Charles Street, Baltimore, MD 21218-2608, e-mail: (jvalana1,vishwanathsindagi,vpatel36)@jhu.edu
			\footnotesize}
		\thanks{ Ilker Hacihaliloglu is with Rutgers, The State University of New Jersey, NJ, USA, e-mail: ilker.hac@rutgers.edu
			\footnotesize}}

\maketitle

\begin{abstract}
Most methods for medical image segmentation use U-Net or its variants as they have been successful
in most of the applications. After a detailed analysis of
these “traditional” encoder-decoder based approaches, we
observed that they perform poorly in detecting smaller
structures and are unable to segment boundary regions
precisely. This issue can be attributed to the increase
in receptive field size as we go deeper into the encoder.
The extra focus on learning high level features causes 
U-Net based approaches to learn less information about
low-level features which are crucial for detecting small
structures. To overcome this issue, we propose using an
overcomplete convolutional architecture where we project
the input image into a higher dimension such that we
constrain the receptive field from increasing in the deep
layers of the network. We design a new architecture for im-
age segmentation- KiU-Net which has two branches: (1) an
overcomplete convolutional network Kite-Net which learns
to capture fine details and accurate edges of the input,
and (2) U-Net which learns high level features. Furthermore,
we also propose KiU-Net 3D which is a 3D convolutional
architecture for volumetric segmentation. We perform a
detailed study of KiU-Net by performing experiments on
five different datasets covering various image modalities. We achieve a good performance
with an additional
benefit of fewer parameters and faster convergence. We also demonstrate that the extensions of KiU-Net
based on residual blocks and dense blocks result in further
performance improvements. Code:  \href{https://github.com/jeya-maria-jose/KiU-Net-pytorch}{https://github.com/jeya-maria-jose/KiU-Net-pytorch}    
\end{abstract}


\section{Introduction}
\label{sec:introduction}
\IEEEPARstart{M}{edical} image segmentation plays a pivotal role in computer-aided diagnosis systems which are helpful in making clinical decisions. Segmenting a region of interest like an organ or lesion from a medical scan is critical as it contains details like the volume, shape and location of the region of interest. Automatic methods proposed for medical image segmentation help in aiding radiologists for making fast and labor-less annotations. Early medical segmentation methods were based on traditional pattern recognition techniques like statistical modeling and edge detection filters. Later, machine learning approaches using hand-crafted features based on the modality and type of segmentation task were developed. Recently, the state of the art methods for medical image segmentation for most modalities like magnetic resonance imaging (MRI), computed tomography (CT) and ultrasound (US) are based on deep learning. As convolutional neural networks (CNNs) extract data-specific features which are rich in quality and effective in representing the image and the region of interest, deep learning reduces the hassle of extracting manual features from the image.

\begin{figure*}[htp!]
	\centering
	\includegraphics[width=0.9\linewidth]{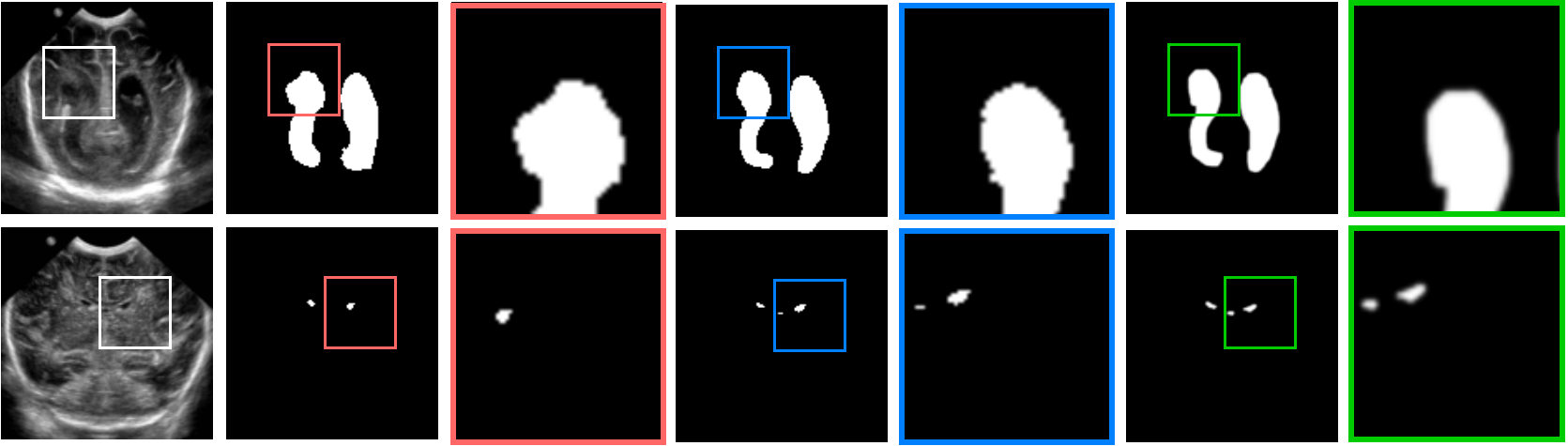}\\
	(a) \hskip50pt (b) \hskip50pt (c) \hskip50pt (d) \hskip50pt (e) \hskip50pt (f) \hskip50pt (g) \hskip50pt  \\
	\caption{ (a) Input B-Mode Ultrasound Image. Predictions from (b) U-Net, (d) KiU-Net
		(ours), (f) Ground Truth. (c),(e) and (g) are the zoomed in patches from (b),(d) and
		(f) respectively. The boxes in the original images correspond to the zoomed in portion
		for the zoomed images. Our method segments small anatomy and edges better than U-Net.}
	
	\label{initvis}
\end{figure*}

Most of the architectures developed for semantic segmentation in both computer vision and medical image analysis are encoder-decoder type convolutional networks. Seg-Net \cite{badrinarayanan2017segnet} was the first such type of network that was widely recognized. In the encoder block of Seg-Net, every convolutional layer is followed by a max-pooling layer which causes the input image to be projected onto a lower dimension similar to an undercomplete auto-encoder. The receptive field size of the filters increases with the depth of the network thereby enabling it to extract high-level features in the deeper layers. The initial layers of the encoder extract low-level information like edges and small anatomical structures while the deeper layers extract high-level information like objects (in the case of vision datasets) and organs/lesions (in the case of medical imaging datasets). A major breakthrough in medical image segmentation was brought by U-Net \cite{ronneberger2015u} where skip connections were introduced between the encoder and decoder to improve the training and quality of the features used in predicting the segmentation. U-Net has became the backbone of almost all the leading methods for medical image segmentation in recent years. Subsequently, many more networks were proposed which built on top of the U-Net architecture. U-Net++ \cite{zhou2018unet++,zhou2019unet++} proposed  using nested and dense skip connection for further reducing the semantic gap between the feature maps of the encoder and decoder. UNet3+ \cite{huang2020unet} proposed using full-scale skip connections where skip connections are made between different scales. 3D U-Net \cite{cciccek20163d} and  V-Net \cite{milletari2016v} were proposed as extensions of U-Net for volumetric segmentation in 3D medical scans. In other extensions of U-Net like Res-UNet \cite{xiao2018weighted} and Dense-UNet \cite{li2018h},   the convolutional blocks in encoder and decoder consisted of residual connections \cite{he2016deep} and dense blocks \cite{huang2017densely} respectively.  It can be noted that all the above extensions of U-Net used the same encoder-decoder architecture and their contributions were either in skip connections, using better convolutional layer connections or in applications.

\begin{figure}[t!]
	\begin{center}
		\centering
		\includegraphics[width=0.15\textwidth,height = 0.15\textwidth]{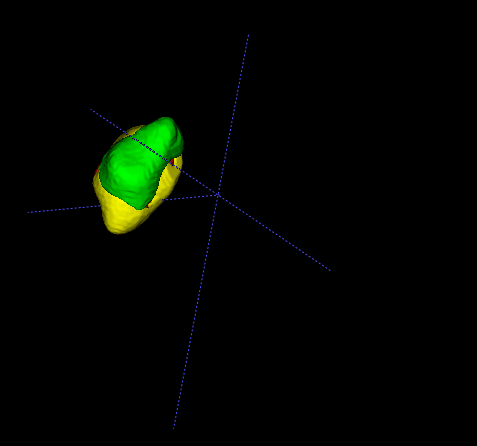}
		\includegraphics[width=0.15\textwidth,height = 0.15\textwidth]{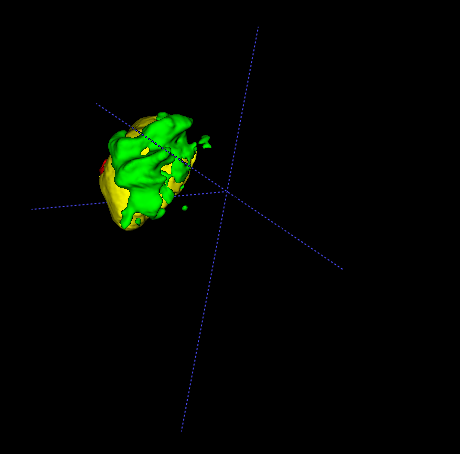}
		\includegraphics[width=0.15\textwidth,height = 0.15\textwidth]{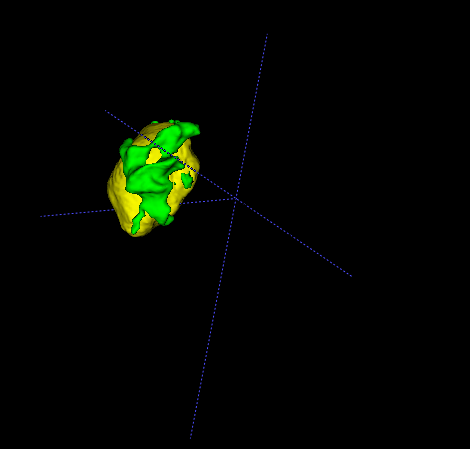} \\
		(a) \hskip65pt (b) \hskip65pt (c)
		\caption{A sample brain tumor segmentation prediction using (a) U-Net 3D, (b) KiU-Net 3D, and (c) Ground-Truth for BraTS dataset. KiU-Net 3D results in better segmentation of fine details when compared to U-Net 3D as it focuses more on low-level information effectively.}
		\label{Fig:first}
	\end{center}
\end{figure}

The main problem with the above family of networks is that they lack focus in extracting features for segmentation of small structures. As the networks are built to be deeper, more high-level features get extracted. Even though the skip connections facilitate transmission of local features to the decoder, from our experiments we observed that they still fail at segmenting small anatomical landmarks with blurred boundaries. Although U-Net and its variants are good at segmenting large structures, they fail when the segmentation masks are small or have noisy boundaries which can be seen in Fig \ref{initvis}. Similarly, in Fig \ref{Fig:first} it can be observed that U-Net 3D \cite{cciccek20163d} fails to extract low-level information like edges and so fails to give a sharp prediction for volumetric segmentation. U-net and its variants belong to undercomplete convolutional architectures which is what causes the network to focus on high-level features. 

To this end, we proposed using overcomplete convolutional architectures for segmentation in \cite{jose2020kiu}. We call our overcomplete architecture Kite-Net (Ki-Net) which transforms the input to higher dimensions (in the spatial sense). Note that Kite-Net does not follow the traditional  encoder-decoder style of architecture, where the inputs are mapped to lower dimensional embeddings (in the spatial sense). Compared to the use  of max-pooling layers in the traditional encoder and upsampling layers in the traditional decoder, Kite-Net has upsampling layers in the encoder and max-pooling layers in the decoder. This ensures  that the receptive field size of filters in the deep layers of the network does not increase like in U-Net thus facilitating Kite-Net to extract fine details of boundaries as well as small structures even in the deeper layers. Although Kite-Net extracts high quality low-level features, the lack of filters extracting  high-level features makes Kite-Net not perform on par with U-Net when the dataset consists of both small and large structure annotations. Hence, we propose a multi-branch network, KiU-Net, where one branch is overcomplete (Ki-Net) and another is undercomplete (U-Net).  Furthermore, we propose to effectively combine the features across the two branches  using a novel cross-residual fusion strategy which results in efficient learning of KiU-Net.

In this journal extension, we make the following contributions: 
\begin{itemize}
	\item We explore the problem of segmenting sharp edges in volumetric segmentation and understand why it is not solved using U-Net 3D and its family of architectures.      
	\item We propose KiU-Net 3D which is an extension of KiU-Net for volumetric segmentation.  Here, we use a 3D convolution-based overcomplete network for efficient extraction of low-level information. 
	\item We propose Res-KiUNet and Dense-KiUNet architectures where we use residual connections and dense blocks respectively for improving the learning of the network.
	\item We evaluate the performance of the proposed architecture  for volumetric and image segmentation across 5 datasets: Brain Tumor Segmentation (BraTS), Liver Tumor Segmentation (LiTS), Gland Segmentation (GlaS), Retinal Images vessel Tree Extraction (RITE) and Brain Anatomy segmentation. These datasets individually correspond to 5 different modalities: ultrasound (US), magnetic resonance imaging (MRI), computed tomography (CT), microscopic and fundus images. With these additional experiments on multiple datasets, we demonstrate   that the proposed method generalizes well to different modalities.
	
\end{itemize}

\section{Related Work}

In this section, we briefly review the deep learning works proposed for medical image segmentation with a focus on capturing fine details. We mainly focus on methods that deal with datasets which we conduct our experiments on.

\subsection{Preserving Local Details:}
There have been a few existing works which focus on improving the performance on local details. In \cite{qu2019improving}, Qu et al. proposes a full resolution convolutional neural network (FullNet) that maintains full resolution feature maps to improve the localization accuracy. In FullNet, there are no max-pooling or downsampling layers which makes the network not increase the receptive field size thus focusing more on boundaries. In \cite{wang2018deepigeos}, Wang et al. proposes an interactive geodesic framework where the resolution-preserving network is proposed by combining user interactions and geodesic distance transforms. It can be noted that although these methods try to solve the problem of segmenting local features better, they do not offer a complete solution. For example, DeepIGeoS requires user interactions and FullNet just tries to have no downsampling layers to reduce the enlargement of receptive field. 

\subsection{Overcomplete Representations:}

In signal processing, overcomplete representations \cite{lewicki2000learning} were first explored for making dictionaries such that the number of basis functions can be more than the number of input signal samples. This enables a higher flexibility for capturing structure in the data. In \cite{lewicki2000learning}, the authors show that overcomplete bases work as a better approximators of any underlying statistical distribution of a data. It has also been widely used for reconstruction of signals under the presence of noise and for source separation in a mixture of signals. It is mainly popular for these tasks because of its greater robustness in the presence of noise when compared to undercomplete representations. For denoising autoencoders \cite{vincent2008extracting}, models with overcomplete hidden layer expression were observed to perform better as they are more useful feature detectors. The authors note that the proposed idea of denoising autoencoders in fact improved the feature detecting ability of an overcomplete fully connected network and it performs better than the standard bottleneck architectures for that task. 

\subsection{Datasets and previous works:}
For brain anatomy segmentation from US scans, methods based on U-Net \cite{martin2018automatic}, PSP-Net \cite{wang2018automatic} have been employed. Also, methods based on confidence based segmentation \cite{valanarasu2020learning} have also been proposed to solve the small anatomy segmentation problem. For gland segmentation, Chen et al. \cite{chen2017dcan} proposed a deep contour-aware network for accurate gland segmentation. Also, Bentaieb et al. \cite{bentaieb2016topology} proposed a topology aware FCN for histology gland segmentation by formulating a new loss to encode geometric priors. There have also been a lot of works that deal with retinal vessel segmentation based on fully convolutional networks \cite{jiang2019automatic}, patch-based methods \cite{feng2017patch} and other multiscale and multilevel deep network configurations \cite{samuel2019multilevel}.     

For brain tumor segmentation for MRI scans,  a lot of methods have been proposed based on 2D U-Net and its variations \cite{weng2019automatic,fang2018three,fridman2018brain,kermi2018deep}.  Many other methods based on Res-Net \cite{he2016deep}, pixel-net \cite{bansal2017pixelnet} and PSP-net \cite{zhao2017pyramid} have been proposed for brain tumor segmentation in \cite{islam2018glioma}. 3D convolution based methods have been proved to be better for segmentation of brain tumor when compared to training 2D convolution networks on individual 2D slices of MRI scans and then  combining them back together to get 3D segmentation. So, 3D U-Net based methods have been proposed in many recent works for brain tumor segmentation. Some of the most notable works for BraTS dataset are as follows: Myronenko et al. \cite{myronenko20183d} where a variational auto-encoder branch is added to reconstruct the input image itself in order to regularize the shared decoder and impose additional constraints on its layers method accurate for brain tumor segmentation. Chen et al. \cite{chen2018s3d} proposes a separable 3D U-Net that use separable 3D convolution thereby reducing computational complexity that the 3D convolutions bring in. Isensee et al. \cite{isensee2018no} showed that how measures like region based training, additional training data and post-processing steps can achieve significant performance boost without any change in architecture. Some of the notable works on LiTS dataset are as follows: Li et al. \cite{li2018h} proposed a densely connected U-Net for liver and tumor segmentation. Zhang et al. \cite{zhang2019light} combined using 2D convolutions and 3D convolutions by using 2D convolutions used at the bottom of the
encoder to decrease the complexity while using  3D convolutions in other layers extract the spatial and temporal information.

\section{Method}

In this section,  we first discuss the issues with the U-Net/U-Net 3D an its family of architectures and motivate why we propose using overcomplete representations. Later, we describe the proposed architectures in detail.

	

\subsubsection{Issues with traditional encoder-decoder networks}

In the dataset that we collected for Brain Anatomy Segmentation from US images, the segmentation masks are heterogeneous in terms of the size of the structures. 
U-Net and its variants   yield relatively good performance for this dataset as seen in \cite{wang2018automatic, valanarasu2020learning}. However, in our experiments we observed that these methods fail to detect tiny structures in most of the cases.  This does not cause much decrement in terms of the overall dice accuracy for the prediction since the datasets predominantly contain images with large structures. However, it is crucial to detect tiny structures with a high precision since it plays an important role in diagnosis.  Furthermore, even for the large structures, U-Net based methods result in erroneous boundaries especially when the boundaries are blurry as seen in Fig \ref{initvis} (b),(c).

In order to clearly illustrate these observations, we evaluated U-Net on the  Brain Anatomy Segmentation from US images  dataset and the results are shown in Fig \ref{initvis}. It can be observed from the bottom row of this figure (Fig \ref{initvis} (b),(c)) that U-Net fails to detect the tiny structures. Further, the first row demonstrates that in the case of large structures,  although U-Net produces an overall good prediction, it is unable to accurately segment out the boundaries.  Additionally, we also made similar observations when U-Net based 3D architecture was used for  volumetric segmentation of lesion. Specifically, the predictions from U-Net are blurred as it fails to segment the surface perfectly especially when the surface of the tumor is not smooth and has a high curvature (see  Fig \ref{Fig:first}(a)). 

To gain a further understanding of why U-Net based models are unable to segment small structures and boundaries accurately, we analyze the network architecture in detail.   In each convolutional block of the encoder, the input set of features to that block get downsampled due to max-pooling layer. This makes sure that the encoder projects the input image to a lower dimension in spatial sense. This combination of convolution layer and max-pooling layer in the encoder causes the receptive field of the filters in deep layers of encoder to increase. With an increased receptive field, the deeper layers focus on high level features and thus are unable to extract features for segmenting small masks or fine edges. The only convolution filters that capture low-level information are the first few  layers. Thus, it can be noted that in a traditional ``encoder-decoder'' architecture, the network is designed such that the number of filters increases as we go deeper in the network \cite{jose2020kiu}.


\begin{figure*}[htbp]
	\centering
	\includegraphics[width=1\linewidth]{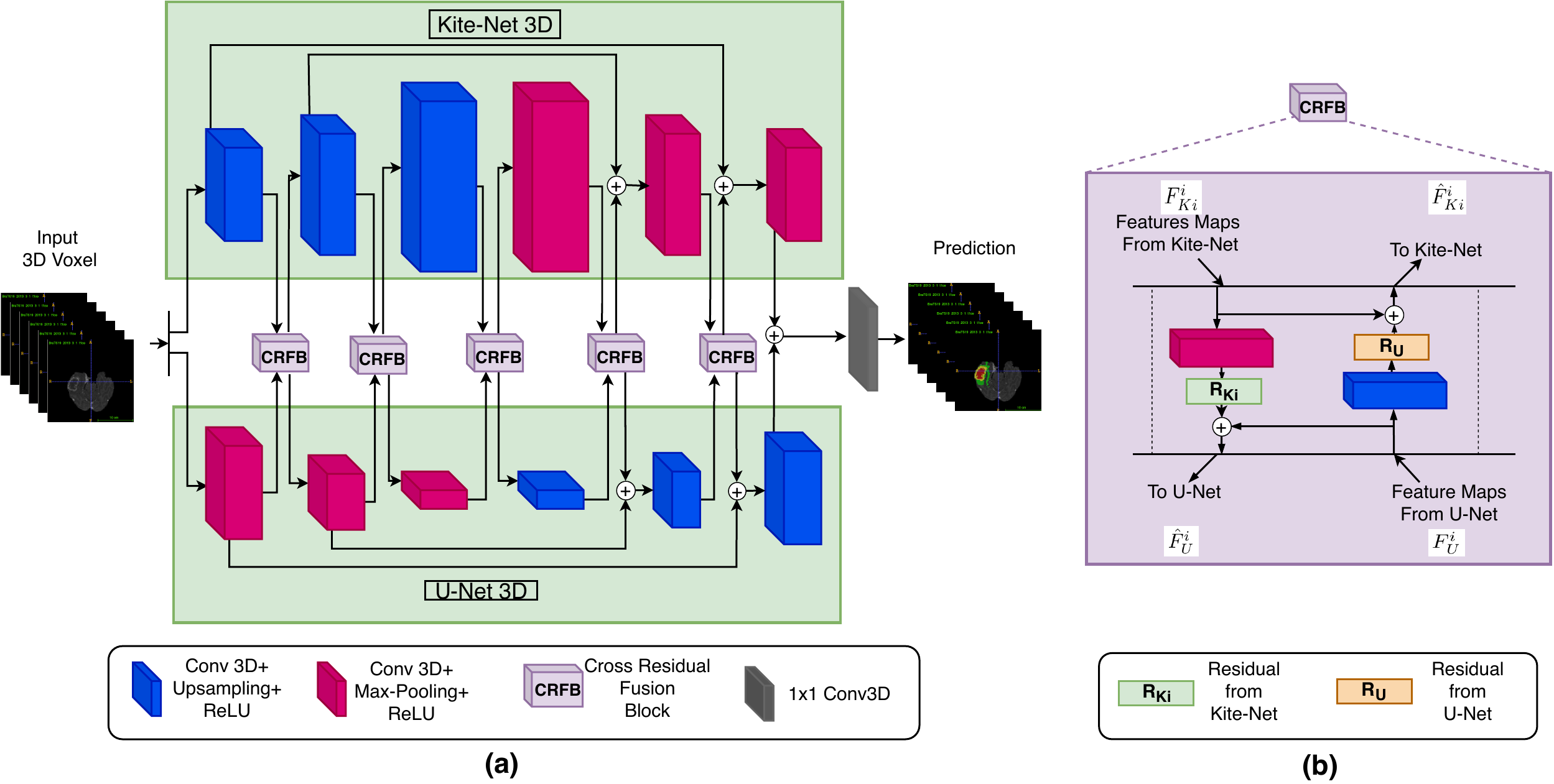}\\
	\caption{(a) Architecture details of KiU-Net 3D for 3D volumetric segmentation. (b) Details of Cross Residual Fusion Block (CRFB) for KiU-Net 3D. In KiU-Net 3D, the input 3D voxel is forwarded to the two branches of KiU-Net 3D: Kite-Net 3D and U-Net 3D which have 3D CRFB blocks connecting them at each level. The feature maps from the last layer of both the branches are added and passed through $1 \times 1$ 3D conv to get the prediction. In CRFB, the residual features of Kite-Net 3D are learned and added to the features of U-Net 3D to forward the complementary features to U-Net and vice-versa.}
	\label{arch2}
\end{figure*}

\subsubsection{Overcomplete Networks}


The idea of overcomplete networks (in the spatial sense) in the convnet-deep learning era has been unexplored. To this end, we propose Kite-Net which is an overcomplete version of U-Net. In Kite-Net, the encoder   projects the input image into a spatially higher dimension. This is achieved by incorporating  bilinear upsampling layers in the encoder. This form of the encoder  constrains the receptive field from increasing like in U-Net as we carefully select the kernel size of the filters and upsampling coefficient such that the deep layers learn to extract fine details and features to segment small masks effectively. Furthermore, in the decoder, each conv block has a conv layer followed by a max-pooling layer. 

To analyze this in detail, let $I$ be the input image, $F_1$ and $F_2$ be the feature maps extracted from first and second conv blocks respectively. Let the initial receptive field of the conv filter be $k \times k$ on the image. In an undercomplete network, the receptive field size change due to max-pooling layer is dependent on two variables: pooling coefficient and stride of the pooling filter. For convenience, the pooling coefficient and stride is both set as 2 in our network. Considering this configuration, the receptive field of conv block 2 (to which $F_1$ is forwarded) on the input image would be $ 2 \times k \times 2 \times k$. Similarly, the receptive field of conv block 3 (to which $F_2$ is forwarded) would be $ 4 \times k \times 4 \times k$. This increase in receptive field can be generalized for an $i^{th}$ layer in an undercomplete network as follows:
\[ RF (w.r.t \; I) =  2^{2*(i-1)} \times k \times k. \]

	

In comparison, the proposed overcomplete network has an upsampling layer with a coefficient 2 in the conv blocks replacing the max-pooling layer. As the upsampling layer actually works  opposite to that of max-pooling layer, the receptive field of conv bock 2 on the input image now would be $ \frac{1}{2} \times k \times \frac{1}{2} \times k$. Similarly, the receptive field of conv block 3  now would be $ \frac{1}{4} \times k \times \frac{1}{4} \times k$. This increase in receptive field can be generalized for $i^{th}$ layer in the overcomplete network as follows:  
\[ RF (w.r.t \; I) =  \left(\frac{1}{2}\right)^{2*(i-1)} \times k \times k. \]

Note that the above calculations are based on a couple of assumptions. We assume that the pooling coefficient and stride are both set as 2 in both overcomplete and undercomplete network. Also, we consider that the receptive field change caused by the conv layer in both undercomplete and overcomplete networks would be the same and do not consider in our calculations. This can be justified as we have maintained the conv kernel size to $3 \times 3$ with stride 1 and padding 1 throughout our network and this setting does not actually affect the receptive as much as max-pooling or upsampling layer does. Similar properties holds true for 3D convolution too. Previous methods like Full-Net try to maintain the resolution of the image similar at each stage. However, upsampling at the encoder constraints the receptive field more as it can be seen from the above calculations.



\subsubsection{Architecture Details of KiU-Net 3D}

With the success of KiU-Net for 2D segmentation \cite{valanarasu2020kiu}, we propose KiU-Net 3D for  volumetric segmentation from 3D medical scans. In the encoder of Kite-Net 3D branch, every conv block has a conv 3D layer followed by a trilinear upsampling layer with coeffecient of two and ReLU activation. In decoder, every conv block has a conv 3D layer followed by a 3D max-pooling layer with coeffecient of two and ReLU activation. Similarly, in  the encoder of U-Net 3D branch, every conv block has a conv 3D layer followed by a 3D max-pooling layer with coefficient of two and ReLU activation. In decoder, every conv block has a conv 3D layer followed by a trilinear upsampling layer with coefficient of two and ReLU activation. We have CRFB block across each layer in KiU-Net 3D similar to KiU-Net. The difference in the CRFB block architecture of KiU-Net 3D is that we have conv 3D layers and trilinear upsampling instead of conv 2D layers and bilinear upsampling like in KiU-Net. The output of both the branches are then added and forwarded to $1 \times 1 \times 1$ conv 3D layer to get the prediction voxel. All the conv layers in our network (except for the last layer) have $3 \times 3 \times 3$ kernel sizes with stride 1 and padding 1. KiU-Net 3D is illustrated in Fig \ref{arch2} (a).

\subsubsection{Cross residual feature block (CRFB)}

In order to further exploit the capacity of the two networks, we propose to combine the features of the two networks at multiple scales through a novel cross residual feature block (CRFB). That is, at each level in the encoder and decoder of KiU-Net 3D, we combine the respective features using a CRFB.  As we know that the features learned by U-Net 3D and Kite-Net 3D are different from each other, this characteristic can be used to further improve the training of the individual networks. So, we try to learn the complementary features from both the networks which will further improve the quality of features learned by the individual networks.

The CRFB block is illustrated in Fig \ref{arch2}.(b). Denoting  the features maps from U-Net 3D as $F_{U}^i$ and  $F_{Ki}^i$ as the feature maps from Kite-Net 3D after $i^{th}$ block in KiU-Net,  the cross-residual features $R_U^i$ and $R_{Ki}^i$ are first extracted using a conv block. The conv block that $F_{Ki}^i$ is forwarded to has a combination of conv 3D layer and max-pooling layer. The conv block that $F_{Ui}^i$ is forwarded through has a combination of conv 3D layer and upsampling layer.  These cross-residual features are then added to the original features $F_{U}^i$ and  $F_{Ki}^i$   to obtain the complementary features $\hat{F}_{U}^i$ and  $\hat{F}_{Ki}^i$,  $\hat{F}_{U}^i = F_{U}^i + R_{Ki}^i$ and $\hat{F}_{Ki}^i = F_{Ki}^i + R_{U}^i$. With this kind of complementary feature extraction from the two networks, we observe a considerable improvement in the segmentation performance of the network.

\section{Experiments}

In this section,  we describe the experimental settings and the datasets  that we use for 2D medical image segmentation and  3D medical volumetric segmentation   to evaluate and compare the proposed  KiU-Net and KiU-Net 3D networks respectively. 

\subsection{KiU-Net}

\subsubsection{Datasets}
\begin{table*}[htbp]
	\centering
	\caption{Performance comparison for 2D Image segmentation with respect to existing methods.}
	\begin{tabular}{c|cc|cc|cc}
		\hline
		Network   & \multicolumn{2}{c|}{Brain US}       & \multicolumn{2}{c|}{GlaS} & \multicolumn{2}{c}{RITE} \\ \hline
		& \multicolumn{1}{c|}{Dice} & p-value & \multicolumn{1}{c|}{Dice}        & p-value     & \multicolumn{1}{c|}{Dice}        & p-value     \\ \cline{2-7} 
		Seg-Net \cite{badrinarayanan2017segnet}  & 0.8279                    & 6.8794e-09        & 0.7861      & 2.8945e-08            & 0.5223      & 1.2856e-11            \\
		U-Net \cite{ronneberger2015u}    & 0.8537                    & 3.8945e-10        & 0.7976      & 5.5374e-07            & 0.5524      & 6.2381e-10            \\
		U-Net++  \cite{zhou2018unet++} & 0.8659                    & 9.0365e-09        & 0.8005      & 4.0036e-07            & 0.5410      & 7.0523e-10            \\
		Full-net \cite{qu2019improving} & 0.8602                    & 6.0234e-08         & 0.7995      & 1.7805e-07             & 0.7054      & 2.6589e-10             \\
		DeepIGeos \cite{wang2018deepigeos} & 0.8354                    & 5.004e-09        & 0.7899      & 4.0245e-08             & 0.5345      & 6.5807e-09            \\
		sSE-UNet \cite{roy2018recalibrating} & 0.8795                    & 8.8852e-08         & 0.8214      &9.5025e-06             & 0.6822      & 1.3665e-11            \\
		CPFNet \cite{feng2020cpfnet}   & 0.8812                    & 6.803e-08         & 0.8237      & 6.3305e-08             & 0.6724      & 5.0255e-09            \\
		KiU-Net   & \textbf{0.8943}                    & -         & \textbf{0.8325}      & -            & \textbf{0.7517}      & -           
	\end{tabular}

	\label{brain}
\end{table*}

\noindent\textit{Brain Anatomy Segmentation (US)}:

 Intraventricular hemorrhage (IVH) which results in the enlargement of brain ventricles is one of the main causes of preterm brain injury. The main imaging modality used for diagnosis of brain disorders in preterm neonates is cranial US because of its safety and cost-effectiveness. Also, absence of septum pellucidum is an important biomarker for septo-optic dysplasia diagnosis. Automatic segmentation of brain ventricles and septum pellucidum from these US scans is essential for accurate diagnosis and prognosis of these ailments. After obtaining institutional review board (IRB) approval, US scans were collected from 20 different premature neonates (age $<$ 1 year). The total number of images collected were 1629 with annotations out of which 1300 were allocated for training and 329 for testing. Before processing, each image was resized to $128 \times 128$  resolution.\\

\noindent\textit{Gland Segmentation (Microscopic)}:
 Accurate segmentation of glands is important to obtain reliable morphological statistics. Histology images in the form of Haematoxylin and Eosin (H$\&$E) stained slides are generally used for gland segmentation \cite{sirinukunwattana2017gland}. Gland Segmentation (GLAS) dataset contains a total of 165 images out of which 85 are taken for training and 80 for testing. We pre-process the images by resizing them to $128 \times 128$ resolution. \\

\noindent\textit{Retinal Nerve Segmentation (Fundus)}: 
Extraction of arteries and veins on retinal fundus images are essential for delineation of morphological attributes of retinal blood vessels, such as length, width, patterns and angles. These attributes are then utilized for the diagnosis and treatment of various ophthalmologic diseases such as diabetes, hypertension, arteriosclerosis and chorodial neovascularization \cite{staal2004ridge}. We use Retinal Images vessel Tree Extraction (RITE) dataset \cite{hu2013automated} which is  a subset of the DRIVE dataset.  RITE dataset contains 40 images split into 20 for training and 20 for testing. We pre-process the images by resizing them to $128 \times 128$ resolution.

\subsubsection{Training and Implementation}

As the purpose of these experiments are to demonstrate the effectiveness of  proposed architecture, we do not use any application specific loss function or metric loss functions in our experiments. We use a binary cross-entropy loss between the prediction and ground truth to train the network.  The cross-entropy loss is defined as follows:

\setlength{\belowdisplayskip}{0pt} \setlength{\belowdisplayshortskip}{0pt}
\setlength{\abovedisplayskip}{0pt} \setlength{\abovedisplayshortskip}{0pt}

\[\mathcal{L}_{CE(p,\hat{p})} = - (\frac{1}{wh} \sum_{x=0}^{w-1}\sum_{y=0}^{h-1}(p(x,y) \log(\hat{p}(x,y)) ) + \]
\[(1-p(x,y))log(1-\hat{p}(x,y))),\]
where $w$ and $h$ are the dimensions of image,  $p(x,y)$ corresponds to the image and  $\hat{p}(x,y)$ denotes the output prediction at a specific pixel location $(x,y)$. We set the batch-size as 1 and learning rate as 0.001. We use Adam optimizer for training.   We use these hyperparameters uniformly across all the experiments. We train our networks for 300 epochs or until convergence depending on the dataset. The experiments were conducted using the Pytorch framework in Python. The experiments were performed using two NVIDIA - RTX 2080 Ti GPUs.

\subsection{KiU-Net 3D}

For volumetric segmentation, we use two widely used public datasets: BraTS challenge dataset and LiTS challenge dataset for our experiments.

\subsubsection{Datasets}

\noindent\textit{Brain Tumor Segmentation (MRI)}: Brain Tumor Segmentation (BraTS) challenge has a curated collection of MRI scans with expert annotations of brain tumor. It contains multimodal MRI scans of confirmed cases of glioblastoma (GBM/HGG) and low grade glioma (LGG). The modalities present in these scans are native (T1), post-contrast T1 (T1ce), T2-weighted (T2) and T2 attenuated inversion recovery (T2-FLAIR) \cite{menze2014multimodal,bakas2017advancing,bakas2018identifying}. The annotations are provided for four classes - enhancing tumor, peritumoral edema and the necrotic and non-enhancing tumor core. We used the 2019 version of the BraTS challenge training data for training our the baselines and our proposed method. It has a total of 335 MRI scans. For quantitative comparisons, we use the validation dataset provided by BraTS 2019 challenge which has 125 scans. For qualitative comparisons, we use the new 33 scans added to the  training set of BraTS 2020 challenge dataset as the validation datasets do not come with ground truth annotations. Each MRI scan contains 155 slices each of dimensions $255 \times 255$.  \\

\noindent\textit{Liver Segmentation (CT)}:  Liver Tumor Segmentation Challenge (LiTS) dataset \cite{bilic2019liver} contains contrast-­enhanced abdominal CT scans along with annotations of liver and liver lesions.  It can be noted that the evaluation for LiTS dataset, we train on the 109 CT scans from the LiTS training dataset and test it on 21 CT scans between 27 and 48 and report the dice accuracy.

\begin{table*}[]
	\caption{Performance comparison for brain tumor volume segmentation in the BraTS dataset with respect to existing methods.}
	\begin{tabular}{c|c|c|c|c|c|c|c|c}
		\hline
		Type & Network & Dice-ET & Dice-WT & Dice-TC & H95-ET & H95-WT & H95-TC &p-value \\ \hline
		& Seg-Net \cite{badrinarayanan2017segnet} &0.4994  &0.7611  &0.6887  &65.6867  &20.3247  &20.0050 & 3.784e-08  \\
		Image & U-Net \cite{ronneberger2015u}&0.5264  &0.8083  &0.7032  &17.5458  &13.9467  &19.2653 &8.568e-10  \\
		& KiU-Net &0.6637  &0.8612 &0.7061  &9.4176  &12.7896  &13.0401 &- \\ \hline
		& Seg-Net 3D \cite{badrinarayanan2017segnet}&0.5599  &0.8062  &0.7073  &10.0037  &10.4584  &10.7513 &1.002e-07 \\
		
		& V-Net  \cite{milletari2016v}& 0.6132 & 0.8318 & 0.7011 & 42.2162 & 20.3525 & 13.0526 &9.665e-08\\
		& Deeper V-Net 3D \cite{milletari2016v}& 0.6251 & 0.8388 & 0.7084 & 39.2542 & 19.4505 & 13.6517 & 6.998e-08\\
		
		& U-Net 3D \cite{cciccek20163d}& 0.6711 & 0.8448 & 0.7059 & 10.2162 & 13.0005 & 15.0856  & 4.225e-06 \\
		Voxel   & Deeper U-Net 3D \cite{cciccek20163d}& 0.6891 & 0.8512 & 0.7294 & 9.8882 & 12.5415 & 12.0452  & 7.568e-10\\
		& Res U-Net 3D \cite{bhalerao2019brain}& 0.6667 & 0.8526 & 0.7091 & 7.270 & 8.5454 & 9.5708  & 9.551e-07\\
		
		& MS U-Net 3D \cite{jesson2017brain, cciccek20163d}& 0.7125 & 0.8652 & 0.7114 & 8.2465 & 9.4205 & 12.6541  & 3.802e-11 \\
		
		& KiU-Net 3D & \textbf{0.7321} & \textbf{0.8760} & \textbf{0.7392} & \textbf{6.3228} & \textbf{8.9424} & \textbf{9.8929} &-\\
		
	\end{tabular}
	\label{brats}
	\centering
\end{table*}

\begin{table*}[]
	\centering
		\caption{Performance comparison for liver and liver lesion segmentation in the LiTS dataset with respect to existing methods.}
	\begin{tabular}{c|c|cc|cc}
		\hline
		Type  & Network     & \multicolumn{2}{c|}{\begin{tabular}[c]{@{}c@{}}Tumor  Segmentation\end{tabular}} & \multicolumn{2}{c}{\begin{tabular}[c]{@{}c@{}}Liver  Segmentation\end{tabular}} \\ \cline{1-6} 
		&             & \multicolumn{1}{c|}{Dice}                        & p-value                         & \multicolumn{1}{c|}{Dice}                        & p-value                         \\ \cline{3-6} 
		& Seg-Net \cite{badrinarayanan2017segnet}     & 0.6452                                           & 2.306e-25                       & 0.7656                                           & 2.998e-22                       \\
		Image & U-Net \cite{ronneberger2015u}       & 0.6950                                           & 6.519e-21                       & 0.7723                                           & 5.665e-16                       \\
		& DeepLab v3+ \cite{chen2017deeplab} & 0.6860                                           & 5.552e-22                       & 0.8570                                           & 3.254e-10                       \\
		& KiU-Net     & 0.7105                                           & -                               & 0.8035                                           & -                               \\ \hline
		& Seg-Net 3D \cite{badrinarayanan2017segnet}  & 0.6822                                           & 3.003e-23                       & 0.8789                                           & 6.028e-25                       \\
		& U-Net 3D \cite{cciccek20163d}    & 0.6912                                           & 9.022e-15                       & 0.9346                                           & 5.005e-10                       \\
		Voxel & Dense-UNet  & 0.5944                                           & 6.556-08                        & 0.9336                                           & 1.985e-11                       \\
		& Li et al.\cite{li2018h}   & 0.7350                                           & 1.887e-09                       & 0.9380                                           & 1.057e-08                       \\
		& KiU-Net 3D  & \textbf{0.7750}                                           & -                               & \textbf{0.9423}                                           & -                              
	\end{tabular}
	\label{lits}
\end{table*}

\subsubsection{Training and Implementation}

For training the KiU-Net 3D, we use the cross entropy loss  between the prediction and the input scan. Since, the scan can be viewed as a 3D voxel,  the cross entropy loss can be formulated as follows:
\setlength{\belowdisplayskip}{2pt} \setlength{\belowdisplayshortskip}{2pt}
\setlength{\abovedisplayskip}{2pt} \setlength{\abovedisplayshortskip}{2pt}
\[\mathcal{L}_{CE(p,\hat{p})} = - \sum_{c=0}^{C-1} \frac{1}{whl} \sum_{z=0}^{l-1} \sum_{x=0}^{w-1}\sum_{y=0}^{h-1}p(z,x,y) \log(\hat{p}(z,x,y) ),\]
where $w$, $h$ are the dimensions of each slice while $l$ is the number of slices in the scan, $p(z,x,y)$ corresponds to the input 3D scan and  $\hat{p}(z,x,y)$ denotes the output prediction at a specific pixel location $(z,x,y)$ and  C corresponds to the number of classes found in the dataset. The above loss formulation suits the multi-class segmentation framework of BraTS dataset where $C=4$. For both the datasets, we use a learning rate of 0.0001 with batch size 1 and use Adam optimizer. We do voxel-wise training (similar to patch-wise training on 2D) for the BraTS dataset with voxel shapes $128 \times 128 \times 50$. The experiments were conducted using the Pytorch framework in Python. The experiments were performed using two NVIDIA - RTX 2080 Ti GPUs.

\section{Results}

In this section, we discuss the results and outcomes of the experiments that we conducted using KiU-Net and KiU-Net 3D. First, we discuss the quantitative evaluations where  the proposed method is compared with other recent approaches using    metrics that are widely used for medical image segmentation. Next, we provide qualitative results where we  visualize sample predictions to analyze why the proposed  method's  performance is superior as compared to  the other approaches.

\subsection{Quantitative Results}

\subsubsection{KiU-Net}

Following existing approaches like \cite{zhou2019unet++}, we use Dice Index (F1-score) for evaluating and comparing the proposed method (KiU-Net) on the medical image segmentation datasets:
\[ Dice = \frac{2TP}{2TP+FP+FN},\]
where TP, FP and FN correspond to the number of pixels that are true positives, false positives and false negatives respectively of the prediction when compared with the ground truth. Table \ref{brain} shows  the results for the experiments on all the 3 datasets for image segmentation. As it can be observed, we compare KiU-Net with some of the widely used backbone architectures for medical image segmentation  like Seg-Net, U-Net, U-Net++, sSE-UNet and CPFNet. We also compare KiU-Net with methods proposed previously for focusing on local details like FullNet and DeepGIoS. The methods under comparison are trained from scratch. Compared to all these, KiU-Net achieves a significant boost in terms of performance. Note that all the other networks were trained from scratch for comparison using the same pipeline as we trained KiU-Net for fair comparison. We also conduct a paired t-test and report the p-value to show the statistical significance of our results. The t-test is conducted between the dice accuracy of our proposed method and all the other baseline methods. Please note that
the p-values are calculated using the dice accuracy for Brain US, GlaS, RITE datasets. We note that all the values are way below 0.05 proving the statistical significance of our results.

\begin{figure}[htbp!]
	\centering
	\includegraphics[width=1\linewidth]{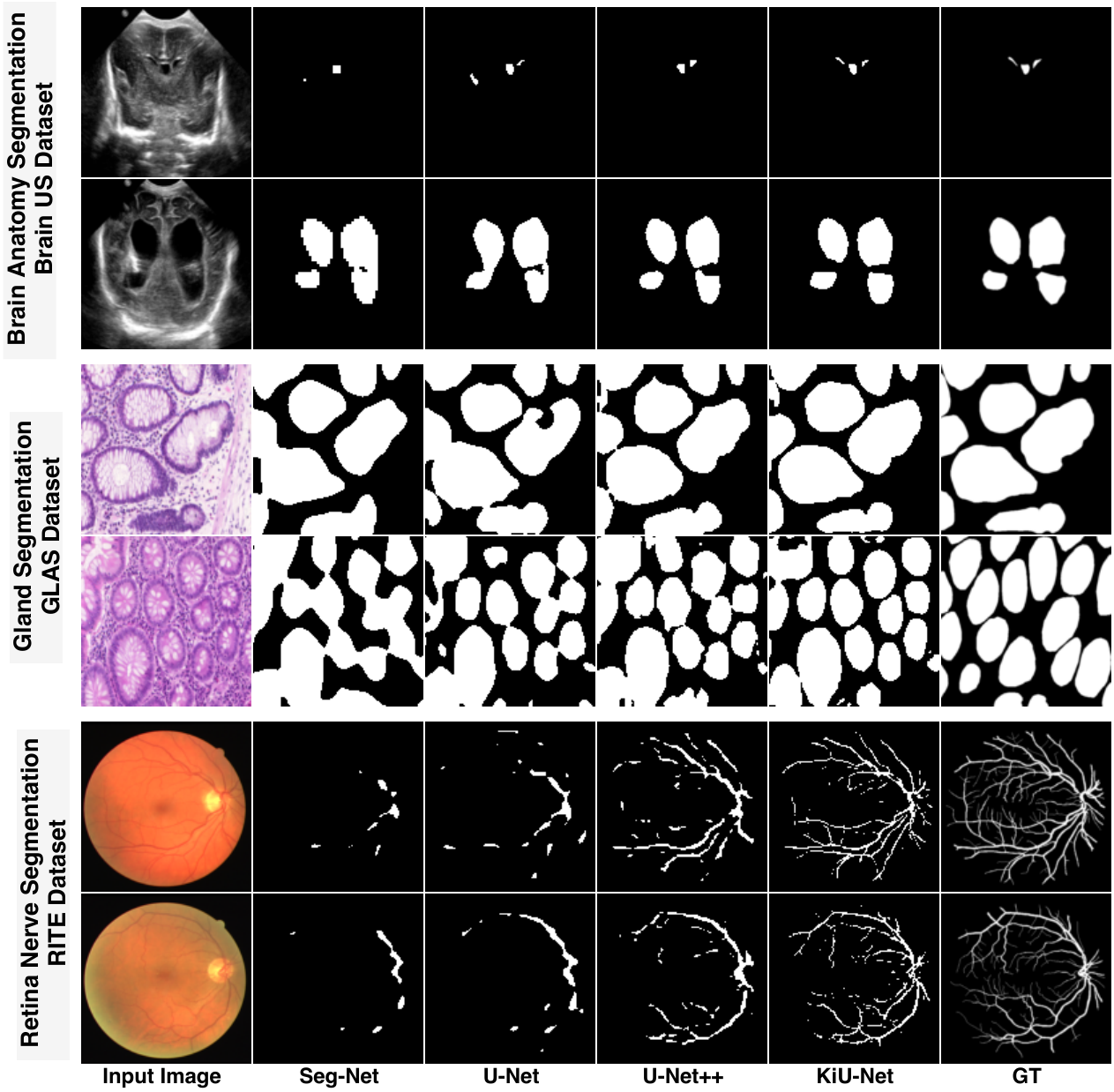}\\

	\caption{Comparison of qualitative results between SegNet, UNet , UNet++ and KiU-Net for Brain anatomy segmentation using the brain US dataset, gland segmentation using the GLAS dataset and retina nerve segmentation using the RITE dataset. }
	\label{res1}
	
\end{figure}

\begin{figure*}[htbp!]
	\centering
	\includegraphics[width=0.9\linewidth]{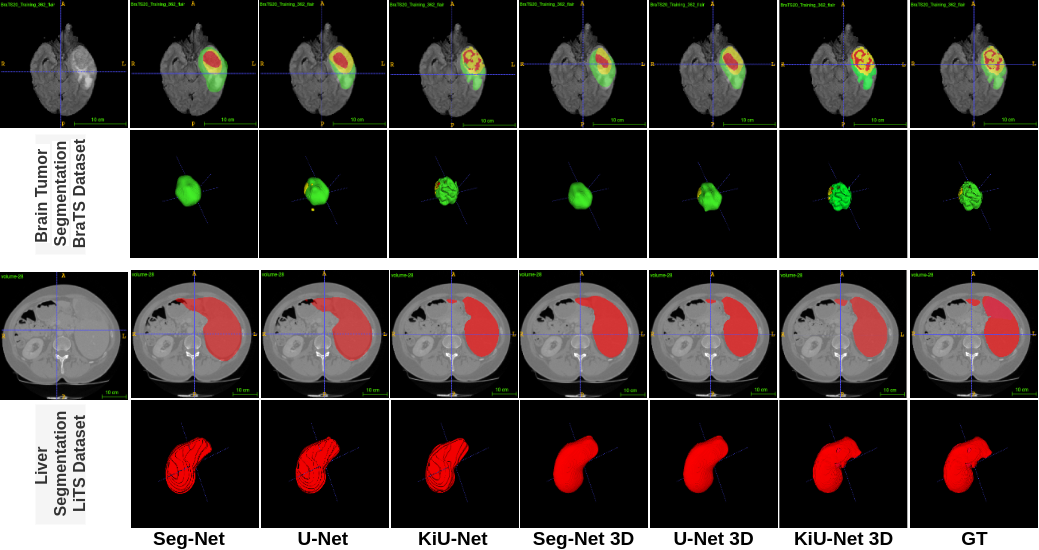}\\

	\caption{Comparison of qualitative results between SegNet, UNet , KiU-Net, SegNet 3D, UNet 3D and KiU-Net 3D for the brain tumor segmentation using the BraTS dataset and the Liver segmentation using the LiTS dataset. The first and third row correspond to the prediction from a 2D slice in the MRI brain scan and abdominal CT scan respectively. The second and third rows visualize the 3D volume segmentation predictions.    }
	\label{res2}
	
\end{figure*} 

\subsubsection{KiU-Net 3D}

For 3D volumetric segmentation, we adopt the performance metrics used in the BraTS challenge and LiTS challenge. For brain tumor segmentation from MRI scans, we report the Dice accuracy of enhancing tumor (ET), whole tumor (WT) and tumor core (TC). We also report the Hausdorff distance for all these three classes. More details about these metrics for brain tumor segmentation can be found in \cite{bakas2018identifying}. Similarly for liver segmentation in LiTS dataset, we report metrics such as Dice score, Jaccard index, volume overlap error (VOE), false negative rate (FNR), false positive rate (FPR),  average symmetric surface distance (ASSD)   and maximum symmetric surface distance (MSD). The details of these metrics can be found in \cite{bilic2019liver}. For both these datasets we compare KiU-Net 3D with U-Net 3D, Seg-Net 3D, V-Net, Res U-Net 3D and MS U-Net 3D. We also compare with some application specific methods like HDense-UNet. Additionally,  we also conduct experiments with  slice-wise training of the scans using 2D versions of all these networks. Tables \ref{brats} and \ref{lits} report the quantitative metrics comparison for BraTS and LiTS dataset respectively.  We observe that KiU-Net 3D outperforms  other methods. The methods under comparison are trained from scratch. Note that we have used the same pipeline for all these experiments for a fair comparison. Since the primary goal of these experiments is to show that KiU-Net can act as a better backbone network architecture, we do not perform  any pre-processing or post-processing   which can  improve the performance further irrespective of the network architecture. We also conduct a paired t-test and report the p-value to show the statistical significance of our results. The t-test is conducted between the dice accuracy of our proposed method and all the other baseline methods. Please note that
the p-values are calculated using the dice accuracy of tumor segmentation in LiTS dataset and dice of whole tumor accuracy for the BraTS dataset. We note that all the values are way below 0.05 proving the statistical significance of our results.

\subsection{Qualitative Results}

\subsubsection{KiU-Net}

Fig \ref{res1} illustrates the predictions of KiU-Net along with Seg-Net, U-Net and U-Net++ for all 3 medical image segmentation datasets we used for evaluation. In the first row, we can observe that KiU-Net is able to segment the small ventricles accurately and  the other ``traditional''  networks fail to do so. Similarly from second, third and fourth rows, we can observe that the predictions of KiU-net are able to segment out the  edges significantly better  when compared to the other networks. In the predictions of RITE dataset which has very low number of images to train, our network performs reasonably well as compared to others  learns, thus demonstrating that low level features for nerve segmentation are used effectively in our network.

\subsubsection{KiU-Net 3D}

Fig \ref{res2} illustrates the predictions of KiU-Net 3D for volumetric segmentation experiments. The first two rows correspond to the results for BraTS dataset and the bottom two rows correspond to the results for LiTS dataset. Note that  the first row and third rows correspond  to segmentation prediction of a single slice of the scan where as the second and fourth row correspond to  the 3D segmentation prediction of the scan. The 3D segmentation results are   visualized using ITKSnap \cite{py06nimg} where each  scan prediction consists of 155 2D images in the  case of BraTS dataset and 48 2D images in the case of LiTS dataset. From  visualizations  of the brain tumor segmentation task, it can be observed that KiU-Net is able to  segment  the surface and edges of the tumor significantly better than any other network and is  closer to the ground truth. For BraTS dataset, the red regions correspond  to tumor core, yellow regions correspond  to non-enhancing tumor and the green regions correspond  to edema. From the second row for BraTS dataset, it can be observed that the tumor surface of the ground truth has sharp edges. While all the other methods smooth out these edges, KiU-Net predicts the sharp edges of the surface of the tumor more precisely.  
While these results are focused on demonstrating the superiority of KiU-Net in segmenting small lesions, the results on the LiTS dataset show that the proposed method is equally effective in segmenting larger regions.   From the bottom two rows of Fig \ref{res2}, we can observe that KiU-Net performs better than other networks in segmenting large masks as well. Based on this, we would like to point out that  even though KiU-Net focuses more on low-level features when compared to UNet or UNet++, its performance is on-par/better as compared to them while segmenting large masks as well. We also observe that performing 2D segmentation on individual 2D images and then combining them to form a 3D scan does not work well. This can be observed from the first 3 images from the last row where the surfaces are not smooth while the ground truth looks smooth.

\section{Discussions}

As we propose a generic architecture as the solution to image and volumetric  segmentation, we believe it is important to study if the network can be further improved using some notable techniques in deep-learning literature. Also, we study other key properties like number of parameters, converge rate and memory requirements of our proposed method in detailed. 

\subsection{Further Improvements}

\begin{figure}[htbp]
	\centering
	\includegraphics[width=1\linewidth]{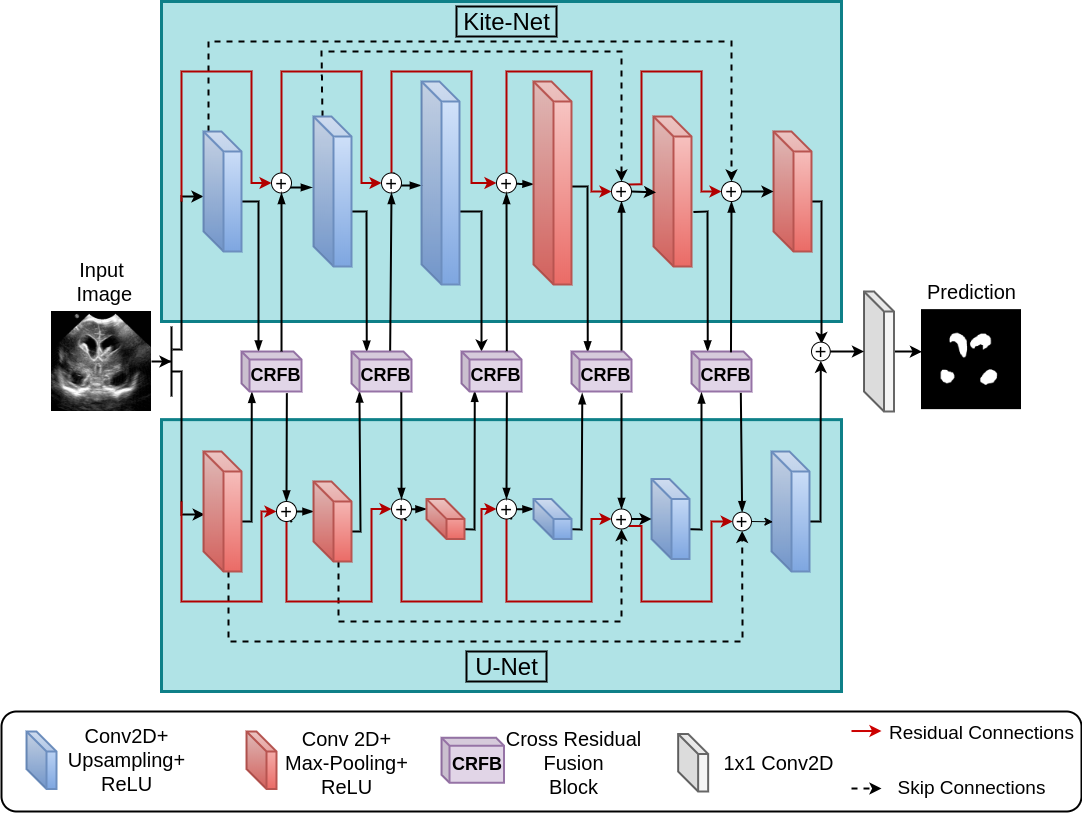}\\

	\caption{Details of Res-KiU-Net architecture. The input image is forwarded to the two branches of Res-KiUNet where each branch has residual connections at each level. The feature maps are added at the last layer and passed through a $1 \times 1$ conv 2D layer to get the prediction.}
	\label{reskiu}
	
\end{figure}

As it is clear from the above discussions that KiU-Net is a good backbone architecture for both image and volumetric segmentation, we  experiment with other variants of KiU-Net which result in further improvements. In this section, we describe these variants and present the detailed results corresponding to each of these variants.

\subsubsection{Res-KiUNet}

\begin{figure}[b]
	\begin{center}
		\centering
		\includegraphics[width=0.09\textwidth,height = 0.09\textwidth]{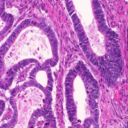}
		\includegraphics[width=0.09\textwidth,height = 0.09\textwidth]{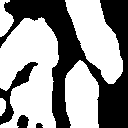}
		\includegraphics[width=0.09\textwidth,height = 0.09\textwidth]{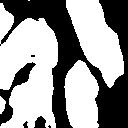}
		\includegraphics[width=0.09\textwidth,height = 0.09\textwidth]{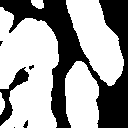}
		\includegraphics[width=0.09\textwidth,height = 0.09\textwidth]{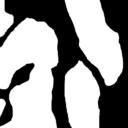} \\
		(a) \hskip37pt (b) \hskip37pt (c) \hskip37pt(d) \hskip37pt (e)
		\caption{(a) Input Image, Prediction using (b) KiU-Net (c) Res-KiUNet (d) Dense KiU-Net (e) Ground Truth for GLAS dataset. It can be observed that the predictions of Res-KiUNet and Dense-KiUNet are better in terms of quality when compared to KiU-Net.    }
		\label{resden}
	\end{center}
\end{figure}

In Res-KiUNet, we employ   residual connections in both the branches of KiU-Net. We use residual learning in every conv block at each level in both the encoder and decoder part of both branches. If $x$ and $y$ are the input and output of each conv block ($F()$) of our network,  the residual connection can be formulated as follows:
\[ y = F(x) + x. \]
We illustrate the architecture details of Res-KiUNet in Fig \ref{reskiu} where the residual connections are denoted using red arrows. Residual connections are helpful in efficient learning of the network since we can propagate gradients to initial layers faster and thus solving the problem of vanishing gradients.

\begin{table}[t]
	\caption{Comparison of performance metrics for Res-KiUNet and Dense KiUNet using the GLAS dataset.}
	\begin{tabular}{
			>{\columncolor[HTML]{FFFFFF}}c 
			>{\columncolor[HTML]{FFFFFF}}c 
			>{\columncolor[HTML]{FFFFFF}}c 
			>{\columncolor[HTML]{FFFFFF}}c }
		\hline
		Performance Metrics & KiU-Net & Res-KiUNet & Dense-KiUNet \\ \hline
		Dice & 0.8325 & 0.8385 & \textbf{0.8431}  \\
		Jaccard & 0.7278 & 0.7303 & \textbf{0.7422} \\ \hline
	\end{tabular}
	\label{fin}
\end{table}

\subsubsection{Dense-KiUNet}

\begin{figure*}[htb]
	\centering
	\includegraphics[width=1\linewidth]{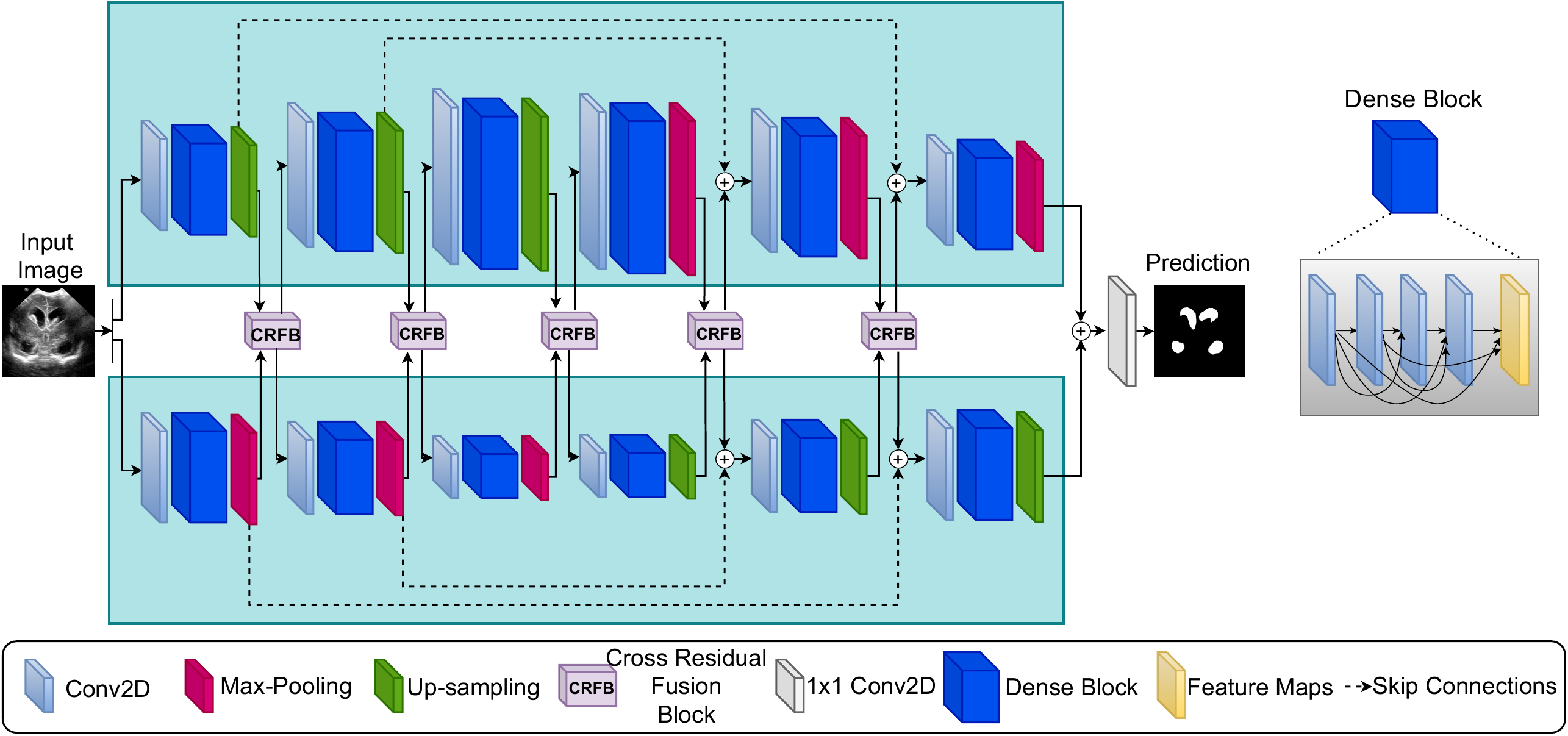}\\

	\caption{Details of Dense-KiU-Net architecture. The input image is forwarded to the two branches of Dense-KiUNet where each branch has dense blocks at each level. The feature maps are added at the last layer and forwarded through a $1 \times 1$ conv 2D layer to get the prediction . In the right side of the figure, dense block architecture has been visualized.  }
	\label{densekiu}
	
\end{figure*}

In Dense-KiUNet, we employ   dense blocks after every conv layer in both the branches. We use a dense block of 4 conv layers where the input consists of $k$ feature maps. Each conv layer outputs $k/4$ feature maps which is concatenated with the input to all the next conv layers.  The output of all these conv layers are then concatenated to obtain $k$ output feature maps. This is added with the input and sent to the next layer in the network. The output of the dense block is forwarded to a max-pooling layer in the encoder of undercomplete branch and in the decoder of overcomplete branch. In  the encoder of overcomplete branch and in the decoder of undercomplete branch, the output of the dense block is forwarded to an upsampling layer. 
Fig \ref{densekiu} illustrates the architecture details of Dense-KiUNet and the dense block we have used.\\

To evaluate both Res-KiUNet and Dense KiUNet, we conduct experiments on the GlaS dataset and report the dice and Jaccard metrics in Table \ref{fin}. Further, we   visualize the results of these experiments in Fig \ref{resden}. It can be observed that Dense-KiUNet and Res-KiUNet provide further improvements in performance as compared to KiU-Net.
\subsection{Number of Parameters}

Seg-Net, U-Net and U-Net++ have a 5 layer deep encoder and decoder. The number of filters in each block of these networks increase gradually as we go deeper in the network. For example, U-Net uses this sequence of filters for its 5 layers - 64, 128, 256, 512 and 1024. Although KiU-Net is a multi-branch network, we limit the complexity of our network by using  fewer     layers and  filters. Specifically,  we use a 3 layer deep network with 32, 64 and 128  respectively as the number of filters. Due to this, the number of parameters in our network is significantly  fewer   as compared to other methods. In Table \ref{param}, we tabulate the number of parameters for KiU-Net and other recent networks and it can be observed   that  KiU-Net has $\sim$10$\times$ lesser parameters as compared to U-Net and $\sim$40$\times$ fewer parameters as compared to SegNet. Further, it is important to note that the other approaches have have resulted in   higher complexity in an attempt to improve the performance,  however, our approach is able to obtain better performance while have significantly fewer parameters.
\begin{table}[htbp]
	\caption{Comparison of number of parameters}
	\scalebox{0.9}{
		\begin{tabular}{
				>{\columncolor[HTML]{FFFFFF}}c 
				>{\columncolor[HTML]{FFFFFF}}c 
				>{\columncolor[HTML]{FFFFFF}}c 
				>{\columncolor[HTML]{FFFFFF}}c 
				>{\columncolor[HTML]{FFFFFF}}c }
			\hline
			Network & Seg-Net \cite{badrinarayanan2017segnet} & U-Net \cite{ronneberger2015u} & U-Net++ \cite{zhou2018unet++} & KiU-Net \\ \hline
			No. of Parameters & 12.5M & 3.1M & 9.0M  & \textbf{0.29M} \\ \hline
			\label{param}
		\end{tabular}
	}
\end{table}

\begin{figure}[t]
	\centering
	\includegraphics[width=0.9\linewidth]{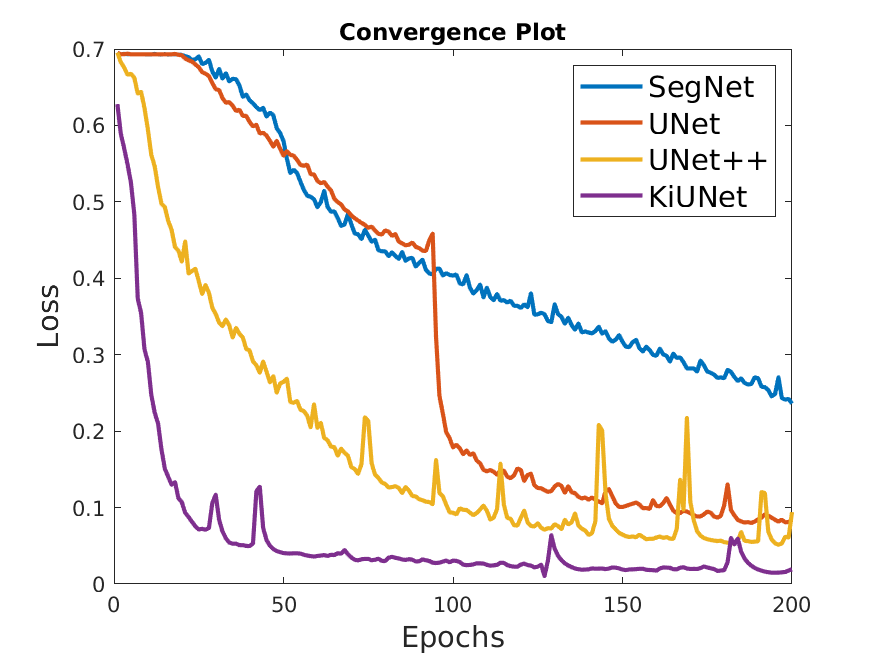}
	\caption{Comparison of convergence of loss function between KiU-Net, UNet++ \cite{zhou2018unet++}, UNet \cite{ronneberger2015u} and SegNet \cite{badrinarayanan2017segnet}. The convergence of KiU-Net is faster when compared to all other methods. }
	\label{convg}
\end{figure}

\subsection{Convergence}

The convergence of loss function is an important characteristic associated with a network.  A faster convergence means is always beneficial as it results in significantly lower training complexity. Fig \ref{convg} compares the convergence trends for Seg-Net, U-Net, U-Net++ and KiU-Net when trained on GLAS dataset. It can be observed that KiU-Net converges faster when compared to other networks. Similar trends were observed while training the network on other datasets as well.

\subsection{Ablation Study}

\begin{figure*}[ht!]
	\begin{center}
		\centering
		\includegraphics[width=0.1\textwidth,height = 0.1\textwidth]{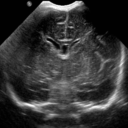}
		\includegraphics[width=0.1\textwidth,height = 0.1\textwidth]{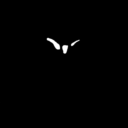}
		\includegraphics[width=0.1\textwidth,height = 0.1\textwidth]{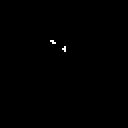}
		\includegraphics[width=0.1\textwidth,height = 0.1\textwidth]{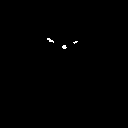}
		\includegraphics[width=0.1\textwidth,height = 0.1\textwidth]{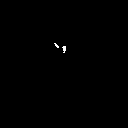}
		\includegraphics[width=0.1\textwidth,height = 0.1\textwidth]{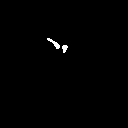}
		\includegraphics[width=0.1\textwidth,height = 0.1\textwidth]{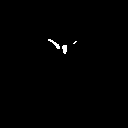}

		\vskip4pt
		\includegraphics[width=0.1\textwidth,height = 0.1\textwidth]{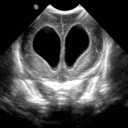}
		\includegraphics[width=0.1\textwidth,height = 0.1\textwidth]{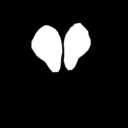}
		\includegraphics[width=0.1\textwidth,height = 0.1\textwidth]{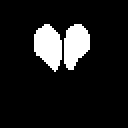}
		\includegraphics[width=0.1\textwidth,height = 0.1\textwidth]{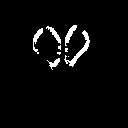}
		\includegraphics[width=0.1\textwidth,height = 0.1\textwidth]{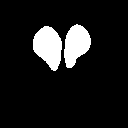}
		\includegraphics[width=0.1\textwidth,height = 0.1\textwidth]{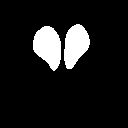}
		\includegraphics[width=0.1\textwidth,height = 0.1\textwidth]{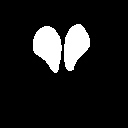}

		
		(a)\hskip42pt(b)\hskip42pt(c)\hskip42pt(d)\hskip42pt(e)\hskip42pt(f)\hskip42pt(g)\\
		\caption{Qualitative results of ablation study on test images. (a) B-Mode input US image. (b) Ground Truth annotation. Prediction of segmentation masks by (c) UC - Under-complete architecture (d) OC - Over-complete architecture (e) UC + SK (under-complete architecture with skip connections) (f) UC + OC with SK (combined architecture with skip connections) (g) KiU-Net (ours). }
		\label{Fig:abl}
	\end{center}
\end{figure*}

We conduct an ablation study to analyze the effectiveness of different blocks   in the  proposed method (KiU-Net). For these experiments, we use the brain anatomy segmentation US dataset. We start with the basic undercomplete (``traditional'') encoder-decoder convolutional architecture (UC) and overcomplete convolutional architecture (OC). Note that these networks do not contain any    skip connections (SK). Next, we add skip connections to the UC and OC baselines. These networks are basically U-Net (UC+skip connections) and Kite-Net (OC+skip connections). We then fuse both these networks by adding the feature map output of both the networks at the end. This is in fact KiU-Net without the CRFB block. Finally, we show the performance of our proposed architecture - KiU-Net.  Table \ref{abl} shows the results of all these ablation experiments. It can be observed that the performance improves   with   addition of each block to the network. Please note that the performances of OC and Kite-Net are  lower because these predictions contain only the edges of the masks and do not contain any high-level information.  Fig \ref{Fig:abl} illustrates the qualitative improvements after adding each major block.

\begin{table}[h]
	\caption{Ablation Study using the Brain US dataset.}
	\scalebox{0.9}{
		\begin{tabular}{
				>{\columncolor[HTML]{FFFFFF}}c 
				>{\columncolor[HTML]{FFFFFF}}c 
				>{\columncolor[HTML]{FFFFFF}}c 
				>{\columncolor[HTML]{FFFFFF}}c 
				>{\columncolor[HTML]{FFFFFF}}c 
				>{\columncolor[HTML]{FFFFFF}}c 
				>{\columncolor[HTML]{FFFFFF}}c }
			\hline
			Metrics & UC & OC & UC+SK & OC+SK & UC+OC+SK & KiU-Net \\ \hline
			Dice & 0.82 & 0.56 & 0.85 & 0.60 & 0.86 & \textbf{0.89} \\
			Jaccard & 0.75 & 0.43 & 0.79 & 0.47 & 0.78 & \textbf{0.83} \\ \hline
		\end{tabular}
	}
	\label{abl}
\end{table}

\begin{table*}[htbp]
	\centering
	\caption{Comparison of performance when training KiU-Net with different depths using the GlaS dataset.}
	\begin{tabular}{c|c|c|c|c|c}
		\hline
		Depth                   & No. of parameters  & Memory                & Training Time (in s)         & Inference Time (in s) & Dice \\ \hline
		2                       & 0.21M                 & 1.25 GB                  &  3.03                  & 0.03               & 0.8111         \\ \hline
		3 & 0.29M  & 5.55 GB  & 30.26  & 0.35               &  0.8325       
	\end{tabular}
	
	\label{tab7}
\end{table*}

\begin{table*}[htbp]
	\centering
	\caption{Comparison of performance when training KiU-Net with different number of filters using the GlaS dataset.}
	\begin{tabular}{c|c|c|c|c|c}
		\hline
		Number of filters                   & No. of parameters  & Memory                & Training Time (in s)        & Inference Time (in s) & Dice \\ \hline
		8,16,32                       & 0.22                 & 2.03 GB                 & 7.02                  & 0.07               &  0.8195         \\ \hline
		16,32,64 & 0.29 M & 5.55 GB   & 30.26  & 0.35                & 0.8325        
	\end{tabular}
	
	\label{tab8}
\end{table*}

\subsection{Memory Required}
We note that the memory required is dependent on the
depth and the number of filters used in the network. So, with a reduced depth and reduced number
of filters, the network’s memory requirements are less while maintaining a decent performance. An
ablation study was performed for the same using GlaS dataset and the results can be observed
in Tables \ref{tab7} and \ref{tab8}. Note that while performing the experiments for reduced depth, the number
of filters were kept the same. While performing the experiments for
different number of filters, the depth was kept as 3 like in the main paper. Note that a generic
U-Net which is 5 layers deep with filters 64,128,256,512 and 1024 gives a F1 score of 0.7976 on
GlaS dataset as reported in the manuscript. It can be seen that even with less computation cost and time, our proposed architecture gives better results than U-Net.

\subsection{Dependence on batch-size}
We conducted experiments where we tried to fit U-Net/U-Net++ to the GPU with the maximum memory capacity (11 GB). The batch size we used for U-Net/U-Net++ was 4 while for KiU-Net we used only a batch size of 1. The comparison in terms of performance on GlaS dataset can be seen in Table \ref{r1q6}. It can be observed that with a higher batch size, the change in performance for U-Net/U-Net++ is negligible while KiU-Net achieves a significant performance boost when compared to the rest. 

\begin{table}[htbp]
	\centering
	\caption{Comparison of performance with different batch sizes for U-Net and U-Net++ using the GlaS dataset.}
	\begin{tabular}{c|c|c|c}
		\hline
		Method    & Batch size & Dice & Jaccard \\ \hline
		U-Net   & 1    & 0.7976 & 0.6763 \\ 
		U-Net & 4    & 0.7998 & 0.6782 \\ \hline
		U-Net++ & 1    & 0.8005 & 0.6893 \\ 
		U-Net++ & 4    & 0.8033 & 0.6910 \\ \hline
		KiU-Net   & 1         & \textbf{0.8325}      & \textbf{0.7278}    
	\end{tabular}
	
	\label{r1q6}
\end{table}

\subsection{Dependence on resolution of image}

In our main experiments, the images were resized to $128 \times 128$ for Brain US, GlaS and RITE datasets. We can also fix the resolution to some higher value like $512 \times 512$ but in that case we would be interpolating some images to a higher resolution as the width and height of most images (in Brain US and GLAS datasets) fall below that. However, to show that we could still utilize KiU-Net for images of higher resolution we resize all images in the RITE dataset to $512 \times 512$ and conduct experiments for the same. In Table \ref{r1q8}, we tabulate the results and compare the difference in performance while conducting the experiments for U-Net and KiU-Net on two different resolutions of the same dataset ($128 \times 128$, $512 \times 512$) for GlaS dataset. It can be observed that there is a slight change in performance for both U-Net and KiU-Net on a higher resolution of the image.

\begin{table}[htbp]
	\centering
	\caption{Comparison of performance with different resolutions of images using the GlaS dataset.}
	\begin{tabular}{c|c|c|c}
		\hline
		Method    & Resolution & Dice & Jaccard \\ \hline
		U-Net   & $128 \times 128$    &79.76  &67.63  \\ 
		U-Net & $512 \times 512$    &80.23  &68.03  \\ \hline
		KiU-Net & $128 \times 128$    &83.25  &72.78 \\ 
		KiU-Net   & $512 \times 512$         & 83.77     & 73.06     
	\end{tabular}
	
	\label{r1q8}
\end{table}

\section{Conclusion}

In this paper, we proposed KiU-Net and KiU-Net 3D for image and volumetric segmentation, respectively. These are two-branch networks consisting  of an undercomplete and   an overcomplete autoencoder.  Our novelty lies in proposing overcomplete convolutional architecture (Kite-Net) for learning small masks and finer details of surfaces and edges more precisely. The two branches are effectively fused using a novel  cross-residual feature fusion method that results effective training.  Further, we experiment with different variants of the proposed method like Res-KiUNet and Dense-KiUNet. We conduct extensive experiments for image and volumetric segmentation on 5 datasets spanning over 5 different modalities. We demonstrate that the proposed  method performs significantly better when compared to the recent segmentation methods. Furthermore, we also show that our network comes with additional benefits such as lower model complexity and  faster convergence.

\bibliography{myref}

\begin{thebibliography}{10}

\bibitem{badrinarayanan2017segnet}
V.~Badrinarayanan, A.~Kendall, and R.~Cipolla, ``Segnet: A deep convolutional
  encoder-decoder architecture for image segmentation,'' {\em IEEE transactions
  on pattern analysis and machine intelligence}, vol.~39, no.~12,
  pp.~2481--2495, 2017.

\bibitem{ronneberger2015u}
O.~Ronneberger, P.~Fischer, and T.~Brox, ``U-net: Convolutional networks for
  biomedical image segmentation,'' in {\em International Conference on Medical
  image computing and computer-assisted intervention}, pp.~234--241, Springer,
  2015.

\bibitem{zhou2018unet++}
Z.~Zhou, M.~M.~R. Siddiquee, N.~Tajbakhsh, and J.~Liang, ``Unet++: A nested
  u-net architecture for medical image segmentation,'' in {\em Deep Learning in
  Medical Image Analysis and Multimodal Learning for Clinical Decision
  Support}, pp.~3--11, Springer, 2018.

\bibitem{zhou2019unet++}
Z.~Zhou, M.~M.~R. Siddiquee, N.~Tajbakhsh, and J.~Liang, ``Unet++: Redesigning
  skip connections to exploit multiscale features in image segmentation,'' {\em
  IEEE transactions on medical imaging}, vol.~39, no.~6, pp.~1856--1867, 2019.

\bibitem{huang2020unet}
H.~Huang, L.~Lin, R.~Tong, H.~Hu, Q.~Zhang, Y.~Iwamoto, X.~Han, Y.-W. Chen, and
  J.~Wu, ``Unet 3+: A full-scale connected unet for medical image
  segmentation,'' in {\em ICASSP 2020-2020 IEEE International Conference on
  Acoustics, Speech and Signal Processing (ICASSP)}, pp.~1055--1059, IEEE,
  2020.

\bibitem{cciccek20163d}
{\"O}.~{\c{C}}i{\c{c}}ek, A.~Abdulkadir, S.~S. Lienkamp, T.~Brox, and
  O.~Ronneberger, ``3d u-net: learning dense volumetric segmentation from
  sparse annotation,'' in {\em International conference on medical image
  computing and computer-assisted intervention}, pp.~424--432, Springer, 2016.

\bibitem{milletari2016v}
F.~Milletari, N.~Navab, and S.-A. Ahmadi, ``V-net: Fully convolutional neural
  networks for volumetric medical image segmentation,'' in {\em 2016 fourth
  international conference on 3D vision (3DV)}, pp.~565--571, IEEE, 2016.

\bibitem{xiao2018weighted}
X.~Xiao, S.~Lian, Z.~Luo, and S.~Li, ``Weighted res-unet for high-quality
  retina vessel segmentation,'' in {\em 2018 9th International Conference on
  Information Technology in Medicine and Education (ITME)}, pp.~327--331, IEEE,
  2018.

\bibitem{li2018h}
X.~Li, H.~Chen, X.~Qi, Q.~Dou, C.-W. Fu, and P.-A. Heng, ``H-denseunet: hybrid
  densely connected unet for liver and tumor segmentation from ct volumes,''
  {\em IEEE transactions on medical imaging}, vol.~37, no.~12, pp.~2663--2674,
  2018.

\bibitem{he2016deep}
K.~He, X.~Zhang, S.~Ren, and J.~Sun, ``Deep residual learning for image
  recognition,'' in {\em Proceedings of the IEEE conference on computer vision
  and pattern recognition}, pp.~770--778, 2016.

\bibitem{huang2017densely}
G.~Huang, Z.~Liu, L.~Van Der~Maaten, and K.~Q. Weinberger, ``Densely connected
  convolutional networks,'' in {\em Proceedings of the IEEE conference on
  computer vision and pattern recognition}, pp.~4700--4708, 2017.

\bibitem{jose2020kiu}
J.~M. Jose, V.~Sindagi, I.~Hacihaliloglu, and V.~M. Patel, ``Kiu-net: Towards
  accurate segmentation of biomedical images using over-complete
  representations,'' {\em arXiv preprint arXiv:2006.04878}, 2020.

\bibitem{qu2019improving}
H.~Qu, Z.~Yan, G.~M. Riedlinger, S.~De, and D.~N. Metaxas, ``Improving
  nuclei/gland instance segmentation in histopathology images by full
  resolution neural network and spatial constrained loss,'' in {\em
  International Conference on Medical Image Computing and Computer-Assisted
  Intervention}, pp.~378--386, Springer, 2019.

\bibitem{wang2018deepigeos}
G.~Wang, M.~A. Zuluaga, W.~Li, R.~Pratt, P.~A. Patel, M.~Aertsen, T.~Doel,
  A.~L. David, J.~Deprest, S.~Ourselin, {\em et~al.}, ``Deepigeos: a deep
  interactive geodesic framework for medical image segmentation,'' {\em IEEE
  transactions on pattern analysis and machine intelligence}, vol.~41, no.~7,
  pp.~1559--1572, 2018.

\bibitem{lewicki2000learning}
M.~S. Lewicki and T.~J. Sejnowski, ``Learning overcomplete representations,''
  {\em Neural computation}, vol.~12, no.~2, pp.~337--365, 2000.

\bibitem{vincent2008extracting}
P.~Vincent, H.~Larochelle, Y.~Bengio, and P.-A. Manzagol, ``Extracting and
  composing robust features with denoising autoencoders,'' in {\em Proceedings
  of the 25th international conference on Machine learning}, pp.~1096--1103,
  2008.

\bibitem{martin2018automatic}
M.~Martin, B.~Sciolla, M.~Sdika, X.~Wang, P.~Quetin, and P.~Delachartre,
  ``Automatic segmentation of the cerebral ventricle in neonates using deep
  learning with 3d reconstructed freehand ultrasound imaging,'' in {\em 2018
  IEEE International Ultrasonics Symposium (IUS)}, pp.~1--4, IEEE, 2018.

\bibitem{wang2018automatic}
P.~Wang, N.~G. Cuccolo, R.~Tyagi, I.~Hacihaliloglu, and V.~M. Patel,
  ``Automatic real-time cnn-based neonatal brain ventricles segmentation,'' in
  {\em 2018 IEEE 15th International Symposium on Biomedical Imaging (ISBI
  2018)}, pp.~716--719, IEEE, 2018.

\bibitem{valanarasu2020learning}
J.~M.~J. Valanarasu, R.~Yasarla, P.~Wang, I.~Hacihaliloglu, and V.~M. Patel,
  ``Learning to segment brain anatomy from 2d ultrasound with less data,'' {\em
  IEEE Journal of Selected Topics in Signal Processing}, 2020.

\bibitem{chen2017dcan}
H.~Chen, X.~Qi, L.~Yu, Q.~Dou, J.~Qin, and P.-A. Heng, ``Dcan: Deep
  contour-aware networks for object instance segmentation from histology
  images,'' {\em Medical image analysis}, vol.~36, pp.~135--146, 2017.

\bibitem{bentaieb2016topology}
A.~BenTaieb and G.~Hamarneh, ``Topology aware fully convolutional networks for
  histology gland segmentation,'' in {\em International Conference on Medical
  Image Computing and Computer-Assisted Intervention}, pp.~460--468, Springer,
  2016.

\bibitem{jiang2019automatic}
Y.~Jiang, H.~Zhang, N.~Tan, and L.~Chen, ``Automatic retinal blood vessel
  segmentation based on fully convolutional neural networks,'' {\em Symmetry},
  vol.~11, no.~9, p.~1112, 2019.

\bibitem{feng2017patch}
Z.~Feng, J.~Yang, and L.~Yao, ``Patch-based fully convolutional neural network
  with skip connections for retinal blood vessel segmentation,'' in {\em 2017
  IEEE International Conference on Image Processing (ICIP)}, pp.~1742--1746,
  IEEE, 2017.

\bibitem{samuel2019multilevel}
P.~M. Samuel and T.~Veeramalai, ``Multilevel and multiscale deep neural network
  for retinal blood vessel segmentation,'' {\em Symmetry}, vol.~11, no.~7,
  p.~946, 2019.

\bibitem{weng2019automatic}
Y.-T. Weng, H.-W. Chan, and T.-Y. Huang, ``Automatic segmentation of brain
  tumor from 3d mr images using segnet, u-net, and psp-net,'' in {\em
  International MICCAI Brainlesion Workshop}, pp.~226--233, Springer, 2019.

\bibitem{fang2018three}
L.~Fang and H.~He, ``Three pathways u-net for brain tumor segmentation,'' in
  {\em Pre-conference proceedings of the 7th medical image computing and
  computer-assisted interventions (MICCAI) BraTS Challenge}, vol.~2018,
  pp.~119--126, 2018.

\bibitem{fridman2018brain}
N.~Fridman, ``Brain tumor detection and segmentation using deep learning u-net
  on multi modal mri,'' in {\em Pre-Conference Proceedings of the 7th MICCAI
  BraTS Challenge}, pp.~135--143, 2018.

\bibitem{kermi2018deep}
A.~Kermi, I.~Mahmoudi, and M.~T. Khadir, ``Deep convolutional neural networks
  using u-net for automatic brain tumor segmentation in multimodal mri
  volumes,'' in {\em International MICCAI Brainlesion Workshop}, pp.~37--48,
  Springer, 2018.

\bibitem{bansal2017pixelnet}
A.~Bansal, X.~Chen, B.~Russell, A.~Gupta, and D.~Ramanan, ``Pixelnet:
  Representation of the pixels, by the pixels, and for the pixels,'' {\em arXiv
  preprint arXiv:1702.06506}, 2017.

\bibitem{zhao2017pyramid}
H.~Zhao, J.~Shi, X.~Qi, X.~Wang, and J.~Jia, ``Pyramid scene parsing network,''
  in {\em Proceedings of the IEEE conference on computer vision and pattern
  recognition}, pp.~2881--2890, 2017.

\bibitem{islam2018glioma}
M.~Islam, V.~J.~M. Jose, and H.~Ren, ``Glioma prognosis: segmentation of the
  tumor and survival prediction using shape, geometric and clinical
  information,'' in {\em International MICCAI Brainlesion Workshop},
  pp.~142--153, Springer, 2018.

\bibitem{myronenko20183d}
A.~Myronenko, ``3d mri brain tumor segmentation using autoencoder
  regularization,'' in {\em International MICCAI Brainlesion Workshop},
  pp.~311--320, Springer, 2018.

\bibitem{chen2018s3d}
W.~Chen, B.~Liu, S.~Peng, J.~Sun, and X.~Qiao, ``S3d-unet: separable 3d u-net
  for brain tumor segmentation,'' in {\em International MICCAI Brainlesion
  Workshop}, pp.~358--368, Springer, 2018.

\bibitem{isensee2018no}
F.~Isensee, P.~Kickingereder, W.~Wick, M.~Bendszus, and K.~H. Maier-Hein, ``No
  new-net,'' in {\em International MICCAI Brainlesion Workshop}, pp.~234--244,
  Springer, 2018.

\bibitem{zhang2019light}
J.~Zhang, Y.~Xie, P.~Zhang, H.~Chen, Y.~Xia, and C.~Shen, ``Light-weight hybrid
  convolutional network for liver tumor segmentation.,'' in {\em IJCAI},
  vol.~19, pp.~4271--4277, 2019.

\bibitem{valanarasu2020kiu}
J.~M.~J. Valanarasu, V.~A. Sindagi, I.~Hacihaliloglu, and V.~M. Patel,
  ``Kiu-net: Towards accurate segmentation of biomedical images using
  over-complete representations,'' in {\em International Conference on Medical
  Image Computing and Computer-Assisted Intervention}, pp.~363--373, Springer,
  2020.

\bibitem{roy2018recalibrating}
A.~G. Roy, N.~Navab, and C.~Wachinger, ``Recalibrating fully convolutional
  networks with spatial and channel “squeeze and excitation” blocks,'' {\em
  IEEE transactions on medical imaging}, vol.~38, no.~2, pp.~540--549, 2018.

\bibitem{feng2020cpfnet}
S.~Feng, H.~Zhao, F.~Shi, X.~Cheng, M.~Wang, Y.~Ma, D.~Xiang, W.~Zhu, and
  X.~Chen, ``Cpfnet: Context pyramid fusion network for medical image
  segmentation,'' {\em IEEE transactions on medical imaging}, vol.~39, no.~10,
  pp.~3008--3018, 2020.

\bibitem{sirinukunwattana2017gland}
K.~Sirinukunwattana, J.~P. Pluim, H.~Chen, X.~Qi, P.-A. Heng, Y.~B. Guo, L.~Y.
  Wang, B.~J. Matuszewski, E.~Bruni, U.~Sanchez, {\em et~al.}, ``Gland
  segmentation in colon histology images: The glas challenge contest,'' {\em
  Medical image analysis}, vol.~35, pp.~489--502, 2017.

\bibitem{staal2004ridge}
J.~Staal, M.~D. Abr{\`a}moff, M.~Niemeijer, M.~A. Viergever, and
  B.~Van~Ginneken, ``Ridge-based vessel segmentation in color images of the
  retina,'' {\em IEEE transactions on medical imaging}, vol.~23, no.~4,
  pp.~501--509, 2004.

\bibitem{hu2013automated}
Q.~Hu, M.~D. Abr{\`a}moff, and M.~K. Garvin, ``Automated separation of binary
  overlapping trees in low-contrast color retinal images,'' in {\em
  International conference on medical image computing and computer-assisted
  intervention}, pp.~436--443, Springer, 2013.

\bibitem{menze2014multimodal}
B.~H. Menze, A.~Jakab, S.~Bauer, J.~Kalpathy-Cramer, K.~Farahani, J.~Kirby,
  Y.~Burren, N.~Porz, J.~Slotboom, R.~Wiest, {\em et~al.}, ``The multimodal
  brain tumor image segmentation benchmark (brats),'' {\em IEEE transactions on
  medical imaging}, vol.~34, no.~10, pp.~1993--2024, 2014.

\bibitem{bakas2017advancing}
S.~Bakas, H.~Akbari, A.~Sotiras, M.~Bilello, M.~Rozycki, J.~S. Kirby, J.~B.
  Freymann, K.~Farahani, and C.~Davatzikos, ``Advancing the cancer genome atlas
  glioma mri collections with expert segmentation labels and radiomic
  features,'' {\em Scientific data}, vol.~4, p.~170117, 2017.

\bibitem{bakas2018identifying}
S.~Bakas, M.~Reyes, A.~Jakab, S.~Bauer, M.~Rempfler, A.~Crimi, R.~T. Shinohara,
  C.~Berger, S.~M. Ha, M.~Rozycki, {\em et~al.}, ``Identifying the best machine
  learning algorithms for brain tumor segmentation, progression assessment, and
  overall survival prediction in the brats challenge,'' {\em arXiv preprint
  arXiv:1811.02629}, 2018.

\bibitem{bilic2019liver}
P.~Bilic, P.~F. Christ, E.~Vorontsov, G.~Chlebus, H.~Chen, Q.~Dou, C.-W. Fu,
  X.~Han, P.-A. Heng, J.~Hesser, {\em et~al.}, ``The liver tumor segmentation
  benchmark (lits),'' {\em arXiv preprint arXiv:1901.04056}, 2019.

\bibitem{bhalerao2019brain}
M.~Bhalerao and S.~Thakur, ``Brain tumor segmentation based on 3d residual
  u-net,'' in {\em International MICCAI Brainlesion Workshop}, pp.~218--225,
  Springer, 2019.

\bibitem{jesson2017brain}
A.~Jesson and T.~Arbel, ``Brain tumor segmentation using a 3d fcn with
  multi-scale loss,'' in {\em International MICCAI Brainlesion Workshop},
  pp.~392--402, Springer, 2017.

\bibitem{chen2017deeplab}
L.-C. Chen, G.~Papandreou, I.~Kokkinos, K.~Murphy, and A.~L. Yuille, ``Deeplab:
  Semantic image segmentation with deep convolutional nets, atrous convolution,
  and fully connected crfs,'' {\em IEEE transactions on pattern analysis and
  machine intelligence}, vol.~40, no.~4, pp.~834--848, 2017.

\bibitem{py06nimg}
P.~A. Yushkevich, J.~Piven, H.~Cody~Hazlett, R.~Gimpel~Smith, S.~Ho, J.~C. Gee,
  and G.~Gerig, ``User-guided {3D} active contour segmentation of anatomical
  structures: Significantly improved efficiency and reliability,'' {\em
  Neuroimage}, vol.~31, no.~3, pp.~1116--1128, 2006.

\end{thebibliography}
\bibliographystyle{ieeetr}
\end{document}